\documentclass[aps,twocolumn,prl,superscriptaddress,showpacs]{revtex4}

\usepackage{graphicx}
\usepackage{dcolumn}
\usepackage{bm}
\usepackage{amsmath}
\usepackage{amssymb}
\usepackage[dvipsnames]{xcolor}
\usepackage[export]{adjustbox}

\begin{document}

\title{Charge localization, frustration relief, and spin-orbit coupling in
U$_3$O$_8$}

\author{Rolando Saniz}
\affiliation{CMT \& NANOlab, Department of Physics, University of Antwerp,
B-2020 Antwerp, Belgium}

\author{Gianguido Baldinozzi}
\affiliation{Universit\'e Paris-Saclay, CentraleSup\'elec, CNRS, SPMS, 91190 Gif-sur-Yvette, France}

\author{Ine Arts}
\affiliation{EMAT \& NANOlab, Department of Physics, University of Antwerp,
B-2020 Antwerp, Belgium}

\author{Dirk Lamoen}
\affiliation{EMAT \& NANOlab, Department of Physics, University of Antwerp,
B-2020 Antwerp, Belgium}

\author{Gregory Leinders}
\affiliation{Belgian Nuclear Research Centre (SCK CEN), Institute for Nuclear Materials
Science, B-2400 Mol, Belgium}

\author{Marc Verwerft}
\affiliation{Belgian Nuclear Research Centre (SCK CEN), Institute for Nuclear Materials
Science, B-2400 Mol, Belgium}
\date{\today}


\begin{abstract}
Research efforts on the low temperature magnetic order and
electronic properties of
U$_3$O$_8$ have been inconclusive so far. Reinterpreting
neutron scattering results, we use group representation theory to show that the ground
state presents collinear out-of-plane magnetic moments, with antiferromagnetic coupling
both in-layer and between layers. Charge localization relieves the initial geometric frustration, generating a slightly distorted honeycomb sublattice with N\'eel order.
We show, furthermore, that
spin-orbit coupling has a giant effect on the conduction band states and band gap value.
Our results allow a reinterpretation of recent optical absorption measurements.
\end{abstract}

\maketitle

U$_3$O$_8$ is a system that stands out among oxide systems because of the expected anisotropic character of magnetic interactions. Indeed, the nuclear structure is layered, charge localization is expected to occur, and magnetic moments are localized onto an almost undistorted underlying triangular lattice. Interactions among layers produce strong anisotropic effects generated by including further neighbors or using a different exchange coupling~\cite{witczak2014,rau2018}.
Therefore, in this quasi-two-dimensional system, display of ordered magnetic configurations such as antiferromagnetic or ferromagnetic is expected. Recently, two papers\cite{miskowiec2021,isbill2022} address the question of the description of the actual magnetic correlations in U$_3$O$_8$ at low temperature. The first paper is experimental, neutron scattering is used to measure magnetic scattering at low temperature, and provides irrefutable evidence of magnetic superlattice reflections below 25~K that correlate with a heat capacity anomaly~\cite{westrum1959} and a magnetic susceptibility peak~\cite{leask1963}. These results provide support for the onset of an antiferromagnetic (AFM) order below $T_N\approx 25$~K. The second of the two papers uses
density functional theory (DFT) to explore a number of configurations, and to determine their respective energies. Unfortunately, the adopted enumerative method is not the result of an exhaustive search, and it cannot provide a robust proof that the configuration with the minimum energy within the considered set is indeed the ground state of this system. 

In this paper we adopt an approach based on group representation theory to settle the matter of the ground state of U$_3$O$_8$. We compare the results of theory with DFT-based calculations and with the experimental evidence, predicting systematic absences of magnetic reflections, and using this feedback to determine the unique model that explains the experimental observations. Then we discuss the implications of this model of anisotropic interactions among magnetic moments, that  can be described by a generalized AFM Heisenberg spin Hamiltonian with in-layer inter-site anisotropic interactions between a ring of six spins, while on-site interactions are responsible for a direct super-exchange mechanism mediated by the $p_x$ orbital of the O atoms bridging the layers. We find that, in the ground state, the frustration of magnetic moments is lifted by charge localization, producing a classical N\'eel-type AFM state where all magnetic moments (intra-layer, but also inter-layer) have a cyclic anti-ferromagnetic coupling.

The precise knowledge of the characteristics of this magnetic ground state is then used to explain the fine features of the band gap.  In this system, spin-orbit coupling (SOC) is of critical importance. We performed a PBE$+U$+SOC to study this effect. We find that the use of SOC strongly affects the states of the conduction band, narrowing the gap by $\sim$38\%, compared to calculations neglecting SOC. We also performed a partially self-consistent quasi-particle $GW$ calculation (QP$GW_0$) including SOC to assess the robust character of this result.
The predicted electronic structure actually explains the salient features of recent optical absorption measurements \cite{he2013}, providing another demonstration of the excellent match between the description of the electronic structure of the ground-state model and experiments.

The room temperature crystal structure of U$_3$O$_8$
was determined with good accuracy by Loopstra in 1964 \cite{loopstra1964}, and electrical conductivity measurements demonstrated the semiconductor nature of U$_3$O$_8$ around the same time~\cite{george1963}. However, no measurements of the band gap were reported at that time. Only during the past decade have experimental researchers started studying pointedly the electronic structure and the magnetic properties of U$_3$O$_8$. There are now several reports of measurements of its band gap using optical spectroscopy techniques~\cite{he2013,enriquez2020,ranasinghe2020}. From the theoretical point of view, various studies using first-principles methods were published~\cite{ranasinghe2020,yun2011,wen2013,brincat2015} before the two already mentioned recent papers on this subject~\cite{miskowiec2021,isbill2022}. 

We performed our calculations with the VASP code~\cite{vasp}, using the projector augmented-wave method (PAW)~\cite{PAW}. The PAWs for oxygen and uranium counted 6 and 14 valence electrons, respectively ($6s^2 6p^6 5f^3 6d^1 7s^2$, for the latter). The exchange-correlation interactions were described within the PBE$+U$ approximation, as applied by Dudarev et al.~\cite{dudarev1997}, using $U_{\rm eff}=U-J=3.96$ eV, as in that seminal work. The energy cut-off was set to 600 eV, energies were converged to within $10^{-6}$ eV, and forces to 0.03 eV/{\AA}. As mentioned above, SOC was taken into account in all our calculations because of its important effect on the electronic properties in other uranium oxide systems: for instance, calculations neglecting SOC yield incorrect U $5f$ occupancies in U$_3$O$_7$, incorrectly predicting a metal instead of a semiconductor character~\cite{leinders2021}. We also performed a $GW$ calculation to benchmark our PBE$+U$ results \cite{gw}. As indicated above, we applied the partially self-consistent quasi-particle approximation to $GW$ (QP$GW_0$), which uses the spectral method to iterate the Green's function and includes the non-diagonal components of the self-energy, as implemented in VASP~\cite{shishkin2007}.

The symmetry group describing the nuclear structure of the room and low temperature phase of U$_3$O$_8$ is $Amm2$ (\#38)~\cite{loopstra1964}. The U atoms are coordinated with seven oxygen atoms, forming edge-sharing UO$_7$ pentagonal bipyramids. The O atoms forming the pentagons and the U atom sit in a plane, forming a stack of dense layers bridged by the apical O atoms of the bipyramids. In this structure, there are two independent U atoms, U1 (Wyckoff position $2a$) and U2 (Wyckoff position $4d$).
Miskowiec et al.~\cite{miskowiec2021} report the appearance of superlattice reflections below $T_N$ that can be generically indexed as $(\frac{1}{2}kl)$. This observation implies a doubling of the $a$ lattice parameter of the room temperature nuclear structure, a direct consequence of a condensation of a magnetic instability at the $\mathbf{q}_Z=(\frac{1}{2}00)$ point at the surface of the Brillouin zone of the non-magnetic parent phase. It is worth of mention that several magnetic reflections of type $(\frac{1}{2}0l)$ are not observed experimentally, a key piece of information for the final choice of the magnetic space group. The magnetic irreducible representations at $\mathbf{q}_Z$  allow the determination of the complete set of compatible magnetic structures that can be used to simulate the scattered intensities of a neutron diffraction experiment. This analysis can also provide the description of the symmetry-adapted phonons compatible with the magnetic instability. 

There are only 4 possible magnetic configurations of magnetic moments, localized at U atom positions, compatible with a condensation at $\mathbf{q}_Z$ (see Supplemental Material for more detailed information).  Each configuration belongs to a different magnetic irreducible representation ($ \Gamma_{mag}= mZ_1\oplus 3 mZ_2\oplus 3 mZ_3\oplus 2 mZ_4$). Two of these configurations involve collinear magnetic moments along the $x$-axis, the direction normal to the dense layers (all magnetic moments have $m_y=m_z=0$):  the magnetic space group induced by the $mZ_1$ irreducible magnetic representation is  $A_{2a}mm2$ (OG\#38.6.270 according to Litvin’s notation~\cite{litvin2008}) and the one induced by $mZ_4$ is the magnetic space group $A_{2a}mm^{\prime}2^{\prime}$ (OG\#38.9.273). The two other configurations involve noncollinear arrangement of the magnetic moments that align in the $yz$-plane (all magnetic moments have $m_x=0$).  The $mZ_2$ magnetic irreducible representation gives the magnetic space group $A_{2a}mm^{\prime}2^{\prime}$ [OG\#38.9.273,
origin shifted by ($\frac{1}{2}$ 0 0)], while $mZ_3$ gives the magnetic space group $A_{2a}mm2$ [OG\#38.6.270, origin shifted by ($\frac{1}{2}$ 0 0)]. These four configurations are the only ones compatible with the experiment and their specific arrangements of the magnetic moments are directly responsible for different intensities of the predicted magnetic reflections. Only the magnetic arrangement induced by the  $mZ_1$ irreducible representation produces the correct set of systematic extinctions for the $(\frac{1}{2}0l)$ magnetic reflections compatible with the experiment.
The corresponding magnetic structure displays interesting features: first of all, the $A_{2a}$ magnetic centring operator requires a stack of two layers of bipyramids UO$_7$ to describe the structure, effectively doubling the lattice along the $x$ direction; interestingly, the magnetic symmetry induced by the $mZ_1$ irreducible representation forbids magnetic moment localization at U1. This correlates favorably with the hypothesis that U1 atoms have $6+$ oxidation state
\cite{kvashnina2013}. Note that the magnetic structure induced by the other irreducible representations would in principle permit magnetic moments on U1.
Moreover, symmetry constrains the magnetic moments of nearest neighbor U2 atoms of a same layer in an AFM configuration without breaking the $A$-centring or changing the $(x,y)$ unit cell dimension and,
moving along $x$ from one layer to the next, the magnetic moments of the the U2 atoms also display an AFM coupling.
Actually, this configuration corresponds to the one with smaller energy among those analyzed in the previous set of DFT calculations~\cite{isbill2022}. Unfortunately, that set does not include the three other configurations that are also predicted for magnetic instabilities at $\mathbf{q}_Z$ and that are candidates to represent excited configurations of the ground state. Our calculations provide slightly higher energies for these alternative models; actually, the $mZ_3$ configuration seems unstable and it converges to a configuration equivalent to the one induced by $mZ_2$. 
To summarize, $A_{2a}mm2$ is the space group of the ground state of U$_3$O$_8$ compatible with the experimental evidence and it corresponds to a stack of two opposite N\'eel-type magnetic states where all U2 magnetic moments in the next layer systematically flip along $x$, and where the U2 atoms within a same layer form a six-spin ring of alternating ordered AFM moments.

\begin{figure}
\includegraphics[width=0.75\hsize]{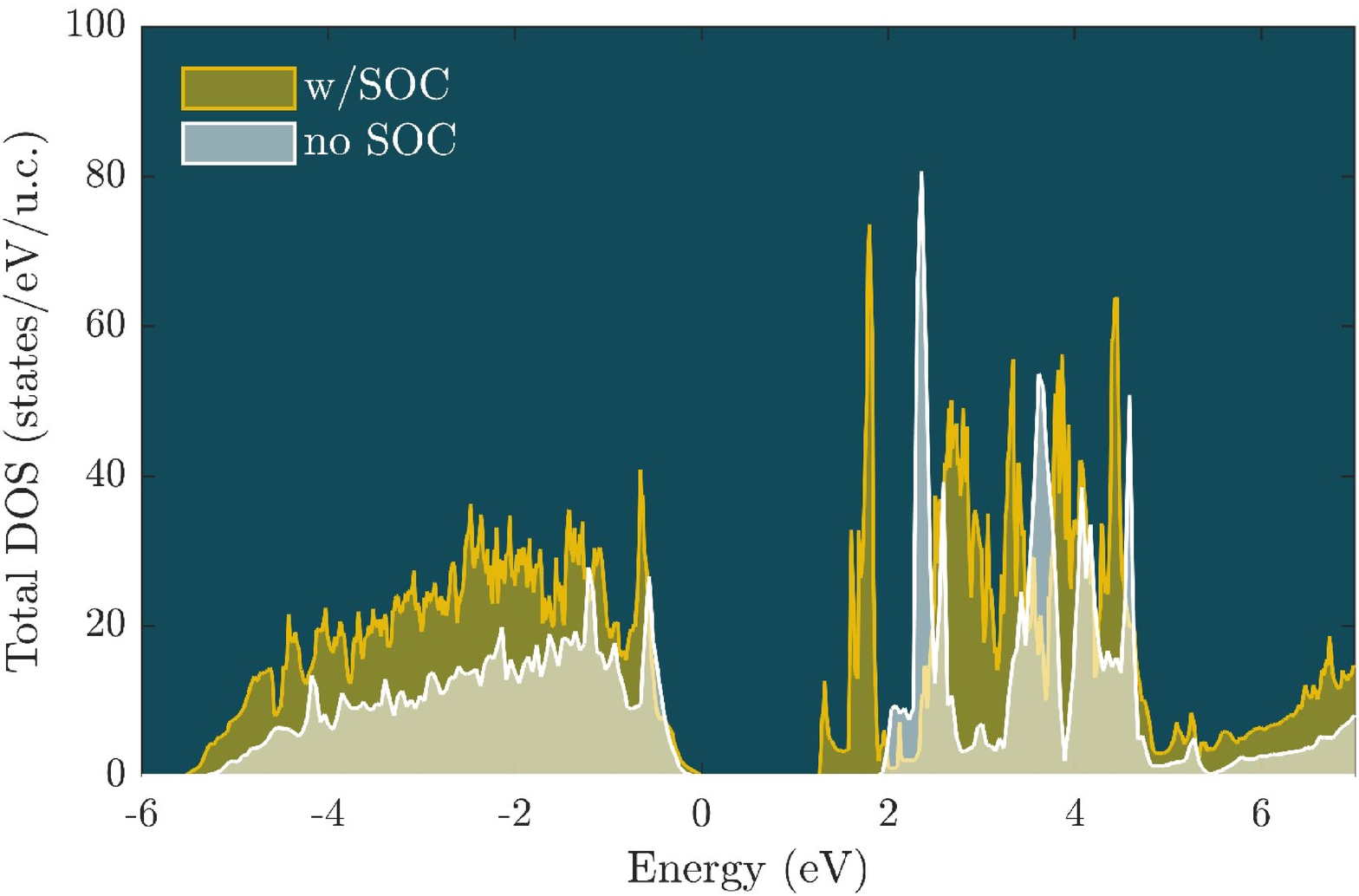}
\includegraphics[width=0.75\hsize]{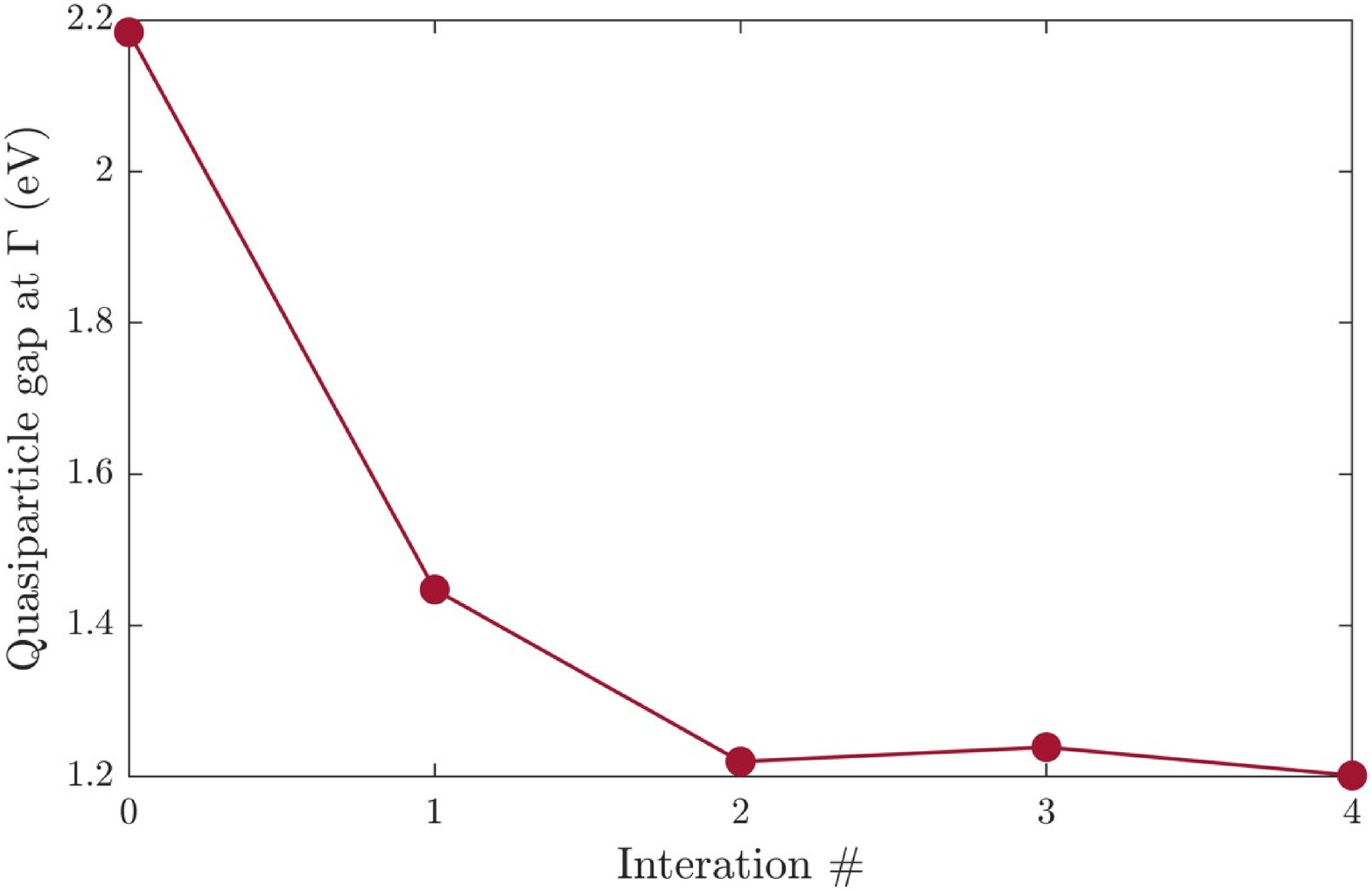}
\caption{\label{fig1} Upper panel:
(Comparison of the total density of states calculations
with and without including spin-orbit coupling. The absence of the latter leads to
considerable energy shifts, affecting binding energies and specially the conduction
band, resulting in a band gap of 2.05 eV.
Lower panel: Convergence of the QP$GW_0$ band gap value at $\Gamma$ with iteration number.
Convergence is typically achieved in four iterations (see www.vasp.at/wiki).}
\end{figure}

The electronic density of states was reported in Ref.~\citenum{isbill2022}, following
a DFT$+U$ approach including SOC. Our results agree qualitatively, and are reported
in the Supplemental Material.
We just note here that our band gap value is 1.27 eV, and the magnetic moments of the U2 atoms
is $\pm0.90$ $\mu_{\rm B}$ (spin and orbital magnetic moments are $\pm2.07$ and $\mp1.17$
$\mu_{\rm B}$, respectively). The small difference with respect to the values reported in
Ref.~\citenum{isbill2022} is probably due to the different Hubbard $U$ parameter value they used.
It is important to recognize that there is a notable cooperative effect between charge localization
and SOC. 
Without charge localization, i.e., without the Hubbard $U$, U$_3$O$_8$ is predicted to be a metal,
with
the manifold of $5f$ states of the U1 and U2 atoms quasi degenerate in energy.
Localization splits the U1 and U2 states around the Fermi level,
allowing the opening of the gap, with the latter states dominating the upper valence band and
the former the lower conduction band.
Therefore, a finite $U$ value is required to drive the system into the AFM phase: from a fundamental point of view, this might have interesting consequences in substituted systems as tuning the Hubbard value can lead to critical properties.
Moreover, minimal seeds of $\pm0.01$ $\mu_{\rm B}$ are sufficient
to obtain the correct AFM order and correct magnetic moments. On the other hand,
SOC acts to strongly lower the cost of occupying the U1 states in the conduction band, thus reducing
significantly the band gap. This is illustrated in the upper panel in Fig.~\ref{fig1}, where
we compare the total density of states of a calculation including SOC
and a calculation neglecting it (with same magnetic moment configuration).
In the latter case
the band gap widens to 2.05 eV, a dramatic change from the 1.27 eV of the SOC calculation.
Such a giant SOC effect on the conduction band has been
reported in other systems, such as hybrid perovskites \cite{even2013}.
Moreover, the amplitude of the magnetic moment of the U2 atoms is
strongly affected by SOC, as they are
estimated to be 1.13 $\mu_{\rm B}$, i.e., $\sim$26 \% larger, when SOC is neglected.

To gauge the band gap value obtained with a PBE+$U$+SOC calculation, we considered a more
accurate approach. To this purpose, we performed a
partially self-consistent quasiparticle QP$GW_0$ calculation (including SOC). In this approach the self-energy (i.e.,
exchange and correlation) is described in a more fundamental way,
yielding band gap values much closer to experiment \cite{shishkin2007}.
Our QP$GW_0$ calculation
gives a band gap value of $\sim$1.20 eV, only $\sim$6\% below the PBE+$U+$SOC result.
We judge that this finding strongly supports the quality and robustness of
our results.
The small difference
 could be an indication that the Hubbard $U$ electronic screening correction value
  used in our calculation is slightly larger
than what is optimally required for U$_3$O$_8$. Be that as it may, a slightly lower $U$ value would have no significant qualitative impact on our results, bringing only small quantitative changes. In the lower panel
of Fig.~\ref{fig1} we show rapid rapid convergence of the QP$QW_0$ band gap value with iteration number.

\begin{figure}
\includegraphics[width=\hsize]{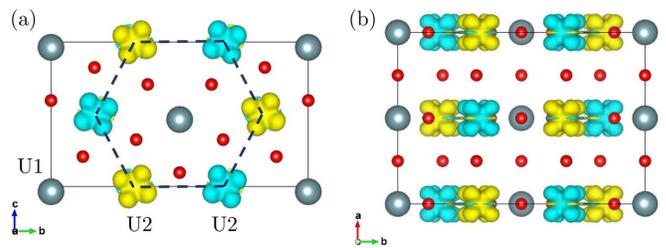}
\caption{\label{fig2} Magnetization density along the direction of the 
$a$-axis. Gold indicates a positive magnetization and cyan a negative one.
(a) Top view of the orthorhombic unit cell. This shows the in-plane
honeycomb N\'eel type AFM order.
(b) Side view of the unit cell, showing the inter-layer AFM coupling.
This strongly
suggest a type of superexchange mechanism mediated by the oxygen atoms between the U2
atoms.}
\end{figure}

Figures~\ref{fig2} show the calculated magnetization density $m_x$.
The gold (cyan) color indicates that the
magnetic moment is in the positive (negative) direction. Figure~\ref{fig2}(a) displays a top view of the unit cell,
clearly showing that the magnetic moments arrange to form a honeycomb lattice with N\'eel type
AFM order. The magnetic moments are quite localized. The magnetization isosurfaces shown
correspond to
a value of only 5\% of the maximum value. 
 The bonding between U2 atoms is asymmetric in
the in-plane directions, suggesting an effective coupling of the magnetic instability
with the $\Gamma_1$ phonons.
Figure~\ref{fig2}(b) shows a side view, exhibiting the interlayer AFM coupling.

\begin{figure}
\includegraphics[width=\hsize]{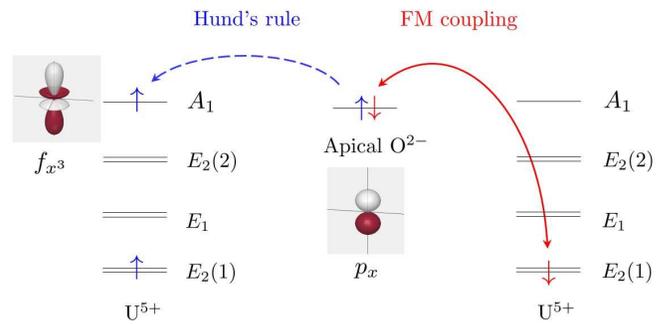}
\caption{\label{fig3} Illustration of the Anderson-Goodenough-Kanamori rule for AFM
superexchange coupling in
the 180$^\circ$ U2-O-U2 configuration.
Thanks to orbital overlap,
an oxygen ion electron 
is virtually excited to a neighboring U2 ion, with a spin complying to Hund's maximum spin rule. The unpaired electron on the O$^-$ ion, which has opposite spin, couples ferromagnetically to the other U2 ion. This is because the $p_x$ orbital is orthogonal
to the $E_2(1)$ states. The effective U2-U2 coupling is AFM.  }
\end{figure}

We posit that the AFM coupling between U2 atoms across layers can be understood in terms of the
180$^\circ$ cation-anion-cation superexchange mechanism advanced by
Anderson, Kanamori, and Goodenough in the 1950s (see Refs.~\citenum{anderson1963,
goodenough1976,kanamori1959}). A precise discussion of the mechanism is complicated by the fact
that spin is no longer a good quantum number because of SOC. However, SOC can be reasonably ignored in
a qualitative discussion of the mechanism. 
First, we point out that the approximate point group of the U2 atoms is $C5v$. Crystal field splits the $f$ orbitals
into states with symmetry corresponding to four irreducible representations
of $C5v$ ($A_1$, $E_1$, $E_2(1)$, and
$E_2(2)$), as shown in Fig.~\ref{fig3}. The lowest lying states belong to $E_2(1)$
and the non-degenerate $A_1$ state has the highest energy, represented by the $f_{x^3}$
spherical harmonic base function.
Using Anderson's terminology \cite{anderson1963}, in the ``ionic configuration'' the
U2 ions have a $5+$ valence state and the bridging O ion a $2-$ valence state. The U2
ions are in their high spin state, while the O ion has zero spin.
In the superexchange mechanism, an electron from the O ion is excited
via a virtual process to one of the U2 ions.
This can occur because the $A_1$ $f_{x^3}$ orbital overlaps with the O $p_x$ orbital, and following Hund's rule, the virtual electron maximizes the total spin virtual configuration of U2. At the same time, the other electron of the O$^{-}$ ion,
which has opposite spin, couples ferromagnetically to other next U$^{5+}$ ion (because
the $p_x$ orbital is orthogonal to the states belonging to $E_2(1)$). Thus, the
resulting U2-U2 coupling across layers is AFM.
The in-plane AFM coupling is more simple and seems to be described by a traditional
anisotropic Heisenberg hamiltonian. Similar anisotropic couplings,
generating a very rich phenomenology,
were seen in other systems \cite{witczak2014,rau2018}.

\begin{figure}
\includegraphics[width=\hsize]{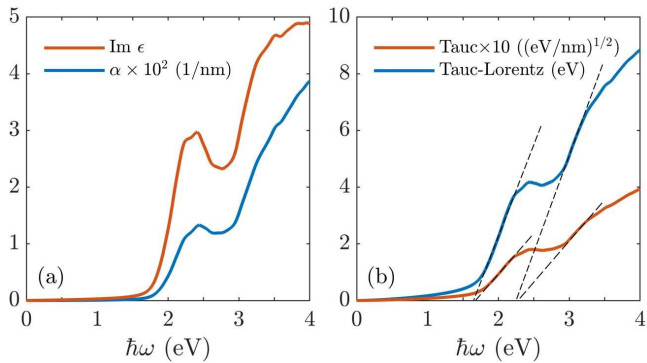}
\caption{\label{fig4} (a) Imaginary part of the dielectric function and absorption coefficient
of U$_3$O$_8$. (b) Tauc and Tauc-Lorentz plots, often used in experiment
to determine indirect band gap values. As explained in the main text, in the present
case the linear extrapolations to the abscissa (represented by the dashed lines) lead to values not directly related to the
band gap in U$_3$O$_8$. We note that our calculated absorption coefficient and Tauc
curves are in very close agreement with the experimental curves in Ref.~\onlinecite{he2013}.}
\end{figure}

Finally, to provide a further comparison with experiment, we calculated the dielectric function and absorption coefficient of U$_3$O$_8$. 
As can be expected,
U$_3$O$_8$ is optically anisotropic, close to uniaxial. We consider, thus, the
isotropic averages (i.e., one third of the trace of the tensors), which can be directly
compared with data from a polycrystalline sample
(see the Supplemental Material for more details).
We focus on the report of He and co-workers \cite{he2013}, as it is more detailed than other
studies \cite{enriquez2020,ranasinghe2020}. We assert that our results explain the reasons for the specific
structural features observed in the absorption spectrum.
In Fig.~\ref{fig4}(a) we present plots of the imaginary part of the dielectric function and of the absorption. 
Absorption starts to develop only as energies approach $\sim$2 eV, clearly indicating
that the joint density of states at lower energies is nearly negligible. After a dip around $\sim$2.7 eV, a second strong increase in
absorption develops around $\sim$3 eV. These features are readily interpreted as arising from U2 to U1 $5f$ transitions
[see the projected densities of states in Fig.~S3(b)]. For comparison with Ref.~\citenum{he2013},
in Fig.~\ref{fig4}(b) we present the plots of two types of Tauc plots. Such plots are often used
in experiment to interpret absorption spectra. Typically, the absorption edge is assumed to indicate the
value of the fundamental band gap.
The plots presented in Ref.~\citenum{he2013} are remarkably close to ours. In that work, the authors
proceed to two linear extrapolations, as shown in
Fig.~\ref{fig4}(b). They then deduce that their sample would contain a mix of
U$_3$O$_8$ and UO$_3$, with the lower absorption edge indicating the band gap of U$_3$O$_8$
and the one above the band gap of UO$_3$ \cite{comment1}.
As our calculation shows, however, the lower absorption edge is not directly related to the band
gap value of U$_3$O$_8$, . Thus, the conclusion in
Ref.~\citenum{he2013} that the band gap of U$_3$O$_8$ falls
between 1.67 and 1.81 eV is due to a
misinterpretation of the Tauc plots they analyze \cite{comment2}.
We stress that the absorption and Tauc plots in Ref.~\citenum{he2013} are in semiquantitative agreement with our results (see Figs. 3, 4, and 5 in Ref.~\citenum{he2013}).
The main difference between our
results and those of He and co-workers is the strength of the second absorption peak, which shows a relatively steeper increase in their results. This maybe be due to the following. Our results indicate
that U$_3$O$_8$ itself gives rise to a second absorption edge starting below
$\sim$3 eV. If the band gap of UO$_3$ is 2.61 eV, as indicated in Ref.~\citenum{he2013},
then the second absorption absorption edge will be steeper in their experiment due to the
combined contribution of U$_3$O$_8$ and UO$_3$.

In this letter, starting from neutron scattering data, we use group representation theory
and first-principles methods to show that the low temperature phase of 
U$_3$O$_8$ has the configuration of a N\'eel state, with simultaneous
in-layer and inter-layer AFM coupling.
The in-layer geometric frustration is relieved by charge localization. Furthermore, SOC has a giant effect on the conduction band states and the band gap value, as it reduces the band gap by almost 0.8 eV compared to a calculation neglecting
SOC. Our calculated optical properties are in close agreement with experiment,
and lead to a basic reinterpretation of the observed spectra.

Financial support for this research was partly provided by the
Energy Transition Fund of the Belgian FPS Economy (Project
SF-CORMOD Spent Fuel CORrosion MODeling). This
work was performed in part using HPC resources from
the VSC (Flemish Supercomputer Center) and the
HPC infrastructure of the University of Antwerp (CalcUA),
both funded by the FWO-Vlaanderen and the Flemish
Government-department EWI.


\begin{thebibliography}{99}
\bibitem{gw} $GW$ calculations are based on an approximation to the self-energy
in terms of the Green's function and the screened Coulomb interaction.
\bibitem{comment1} A caveat is appropriate here: the Tauc approximation in principle assumes
parabolic bands, while in the case of U$_3$O$_8$ the bands around the conduction band minimum are
flat [see Fig.~S3(a)].
\end{thebibliography}
\end{document}


\title{Supplemental Material for \\ Charge localization, frustration relief, and spin-oribt coupling in U$_3$O$_8$}

\author{Rolando Saniz}
\affiliation{CMT \& NANOlab, Department of Physics, University of Antwerp,
B-2020 Antwerp, Belgium}

\author{Gianguido Baldinozzi}
\affiliation{Universit\'e Paris-Saclay, CentraleSup\'elec, CNRS, SPMS, 91190 Gif-sur-Yvette, France}

\author{Ine Arts}
\affiliation{EMAT \& NANOlab, Department of Physics, University of Antwerp,
B-2020 Antwerp, Belgium}

\author{Dirk Lamoen}
\affiliation{EMAT \& NANOlab, Department of Physics, University of Antwerp,
B-2020 Antwerp, Belgium}

\author{Gregory Leinders}
\affiliation{Belgian Nuclear Research Centre (SCK CEN), Institute for Nuclear Materials
Science, B-2400 Mol, Belgium}

\author{Marc Verwerft}
\affiliation{Belgian Nuclear Research Centre (SCK CEN), Institute for Nuclear Materials
Science, B-2400 Mol, Belgium}
\date{\today}

\maketitle

\subsection{I. Unit cell}

\begin{figure}[!h]
\begin{center}
\includegraphics[width=0.8\hsize]{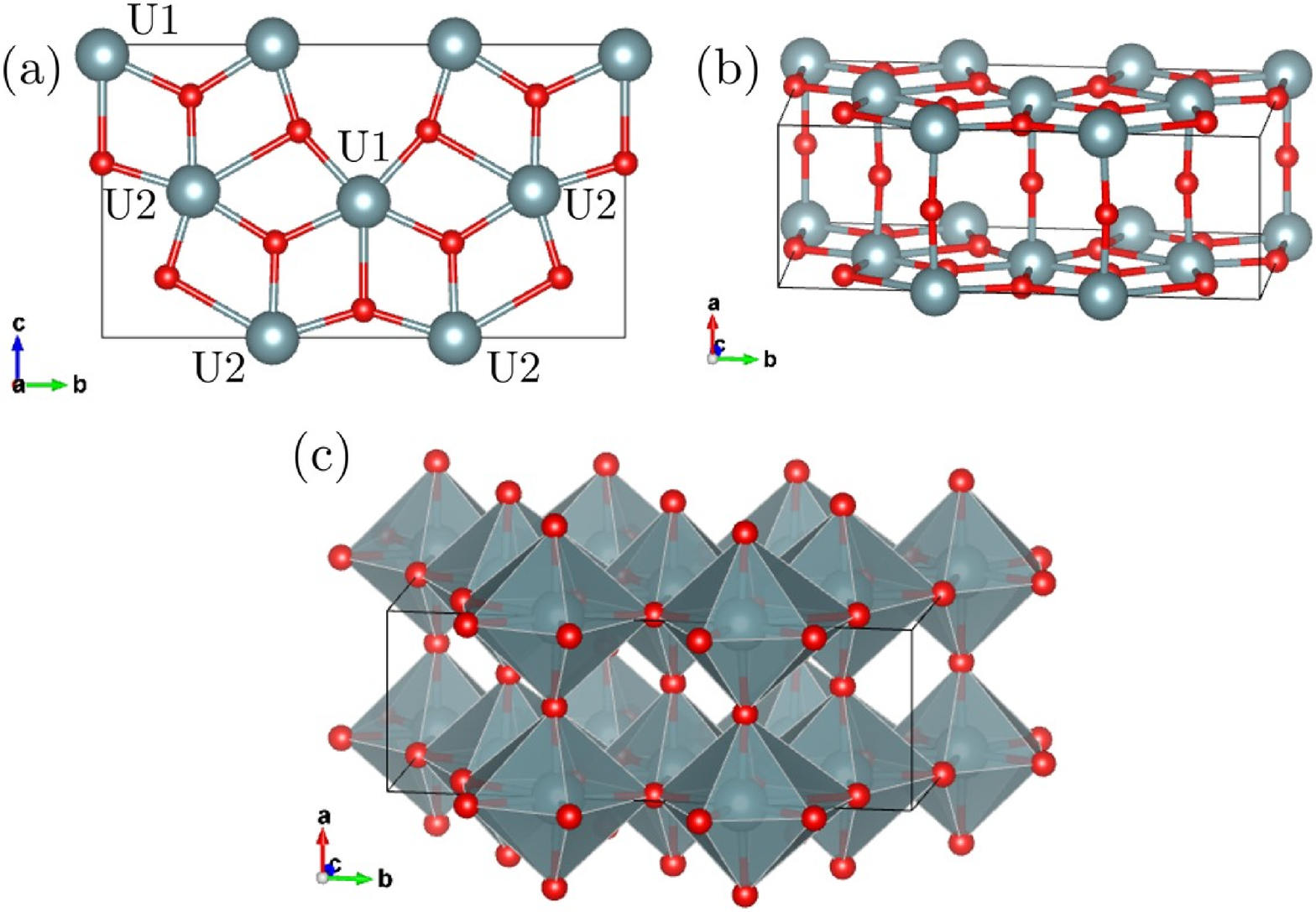} 
\end{center}
\caption{\label{fig1} Low temperature structure of U$_3$O$_8$. 
(a) Top view of the $Amm2$ orthorhombic cell. The U atom labels are for reference in the text.
(b) Perspective view of the same cell, showing the layered character of the structure.
(c) The oxygen pentagonal bipyramids coordinating the U atoms.}
\end{figure}

\clearpage

\subsection{II. Derivation of the magnetic subgroups of the U$_3$O$_8$ $Amm2$ structure}

Magnetic phase transitions involving ordering of localized magnetic moments break the symmetry of the non-magnetic parent phase. The magnetic ordering is then described using an eigenvector basis because, for any ordered property expressed as propagation vectors of the Brillouin zone of the crystal, Wigner’s theorem warrants that eigenvectors transform as irreducible representations. In the current case, the simplest propagation vector compatible with the observed reflections is the
$Z$ point $(\frac{1}{2} 0 0)$ of the centred orthorhombic Brillouin zone of the $Amm2$ parent structure. The magnetic representation of the magnetic moments of the U atoms decomposes onto the following unidimensional irreducible representations of the parent group:
\begin{equation}
mZ_1\oplus 3\,mZ_2\oplus 3\,mZ_3\oplus 2\,mZ_4 \nonumber
\end{equation}

\noindent
Each one of these four irreducible representations gives a compatible magnetic space group describing the magnetic order in the daughter phase:

\begin{table}[!h]
\begin{ruledtabular}
\begin{tabular}{lll}
 Magnetic Irrep & Magnetic Space Group & Comments \\
\hline
 $mZ_1$    & $A_{2a}mm2$ OG \#38.6.270 & AFM, collinear, U1 non-magnetic  \\ [2pt]
 $mZ_2$    & $A_{2a}mm'2$ OG \#38.9.273 & Planar, non-collinear  \\ [2pt]
 $mZ_3$    & $A_{2a}mm2$ OG \#38.6.270 & Planar, non-collinear  \\ [2pt]
 $mZ_4$    & $A_{2a}mm'2$ OG \#38.9.273 & FM intralayer, AFM interlayer, collinear  \\ [2pt]
\end{tabular}
\end{ruledtabular}
\end{table}

\begin{figure}[!h]
\begin{center}
\includegraphics[width=\hsize]{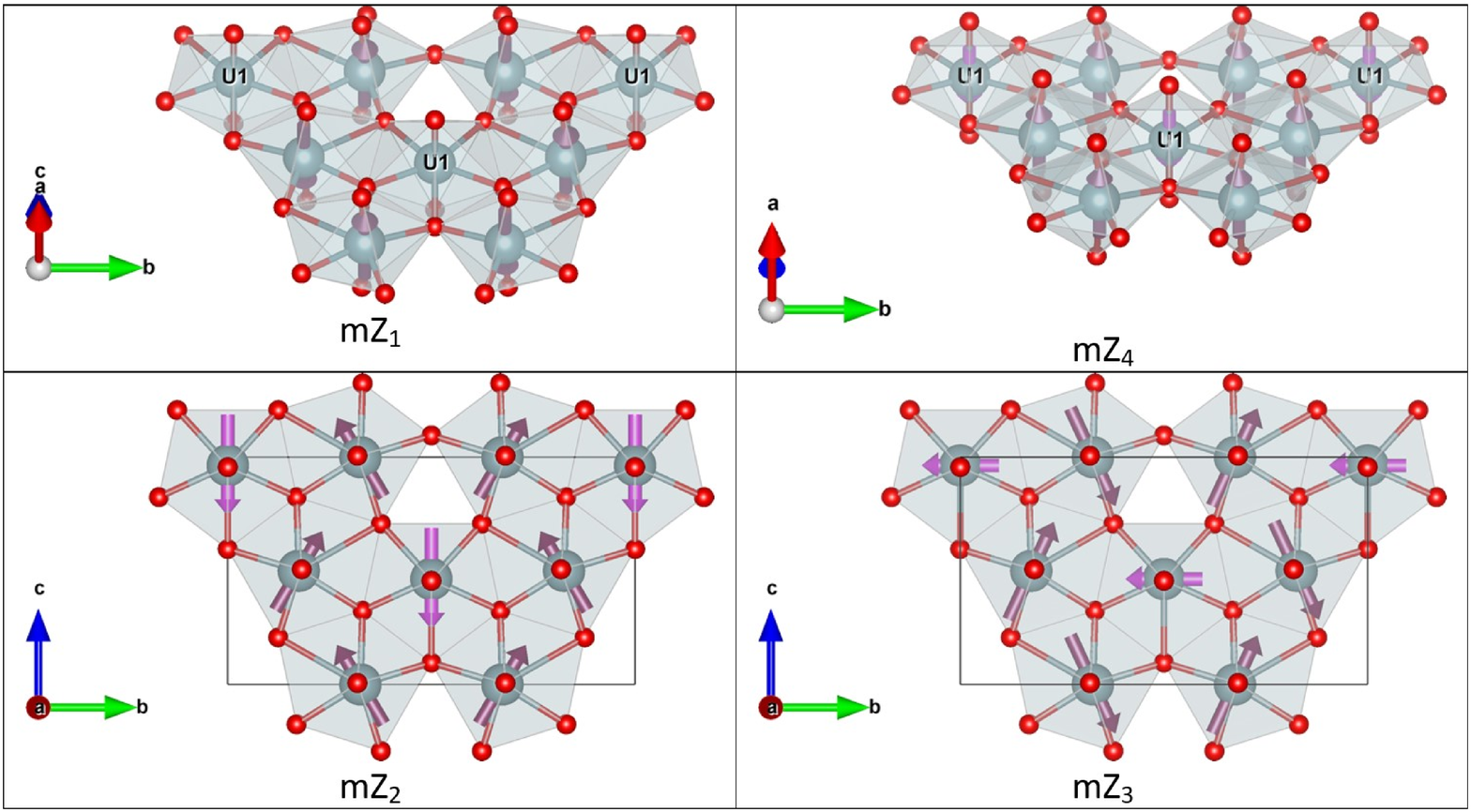} 
\end{center}
\caption{\label{fig2} The schematic figures below display the characteristic ordering of the magnetic moments of U atoms within each one of the four possible structures. Only one of the two layers composing the magnetic structure is represented. In the next layer magnetic moments are always flipped.}  
\end{figure}

\clearpage

List of the low {\bf q} magnetic reflections predicted by the $mZ_1$ AFM model. The reflections indexes refer to the magnetic lattice with the a lattice parameter doubled with symmetry
$A_{2a}mm2$ (38.6.270) in the OG symbol convention \cite{litvin2008}.

\begin{table}[!h]
\begin{ruledtabular}
\begin{tabular}{cccc}
 HKL & $q$ (\AA$^{-1}$) & Model prediction & Miskowiec \cite{miskowiec2021} \\
\hline
(100)    & 0.76 & extinct  & \\ [2pt]
(110)    & 0.92 & extinct  & \\ [2pt]
(101)    & 1.20 & extinct  & \\ [2pt]
(120)    & 1.29 & extinct  & \\ [2pt]
(111)    & 1.31 & ``strong'' & observed \\ [2pt]
(121)    & 1.60 & extinct &  \\ [2pt]
(130)    & 1.75 & extinct &  \\ [2pt]
(131)    & 1.98 & very weak & not observed \\ [2pt]
(102)    & 2.02 & extinct &  \\ [2pt]
(112)    & 2.09 & extinct &  \\ [2pt]
(140)    & 2.23 & extinct &  \\ [2pt]
(300)    & 2.27 & extinct &  \\ [2pt]
(122)    & 2.28 & very weak & observed \\ [2pt]
(310)    & 2.33 & extinct  &
\end{tabular}
\end{ruledtabular}
\end{table}

\clearpage

\subsection{III. Electronic structure}

\begin{figure}[!h]
\includegraphics[width=0.7\hsize]{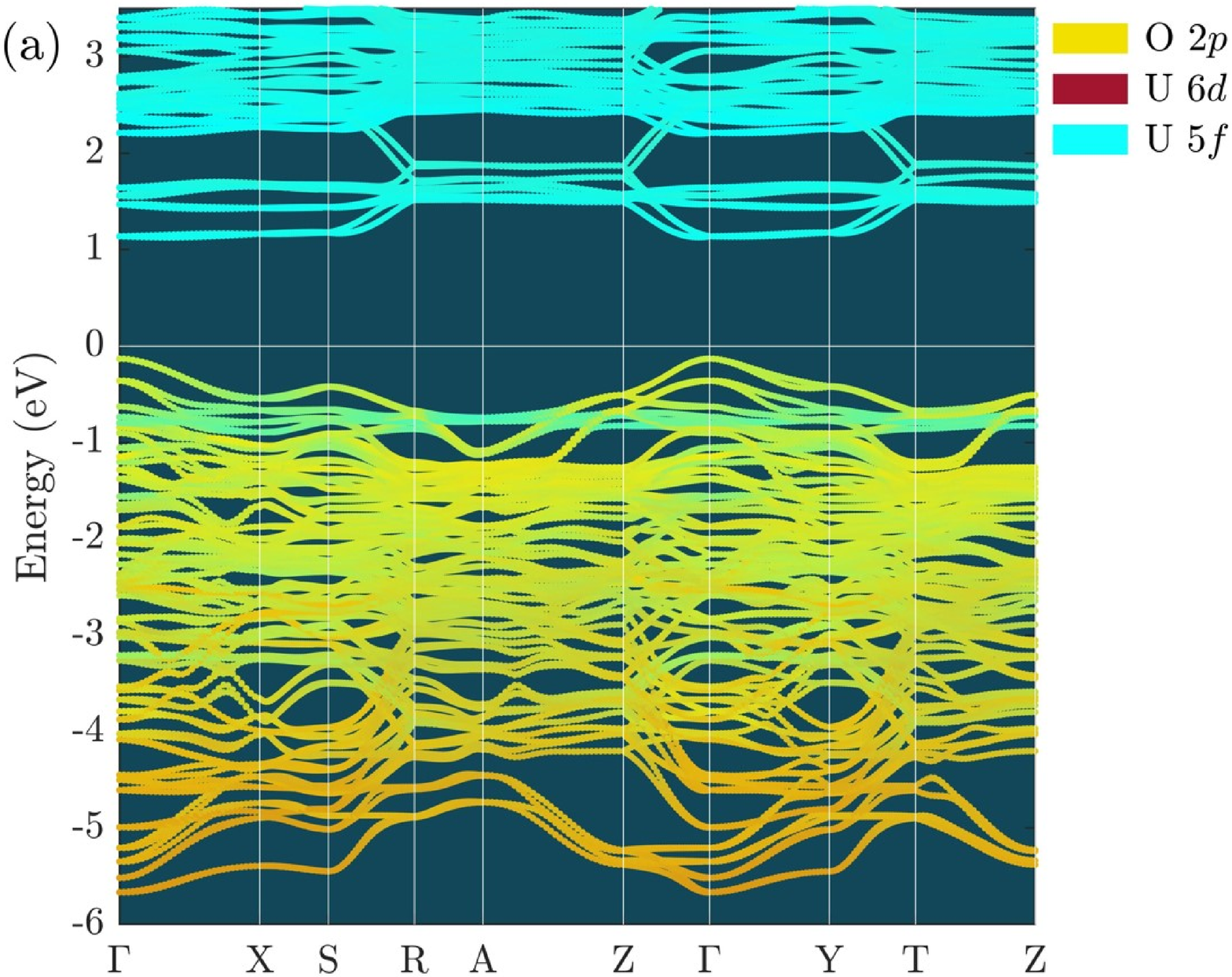}
\includegraphics[width=0.7\hsize]{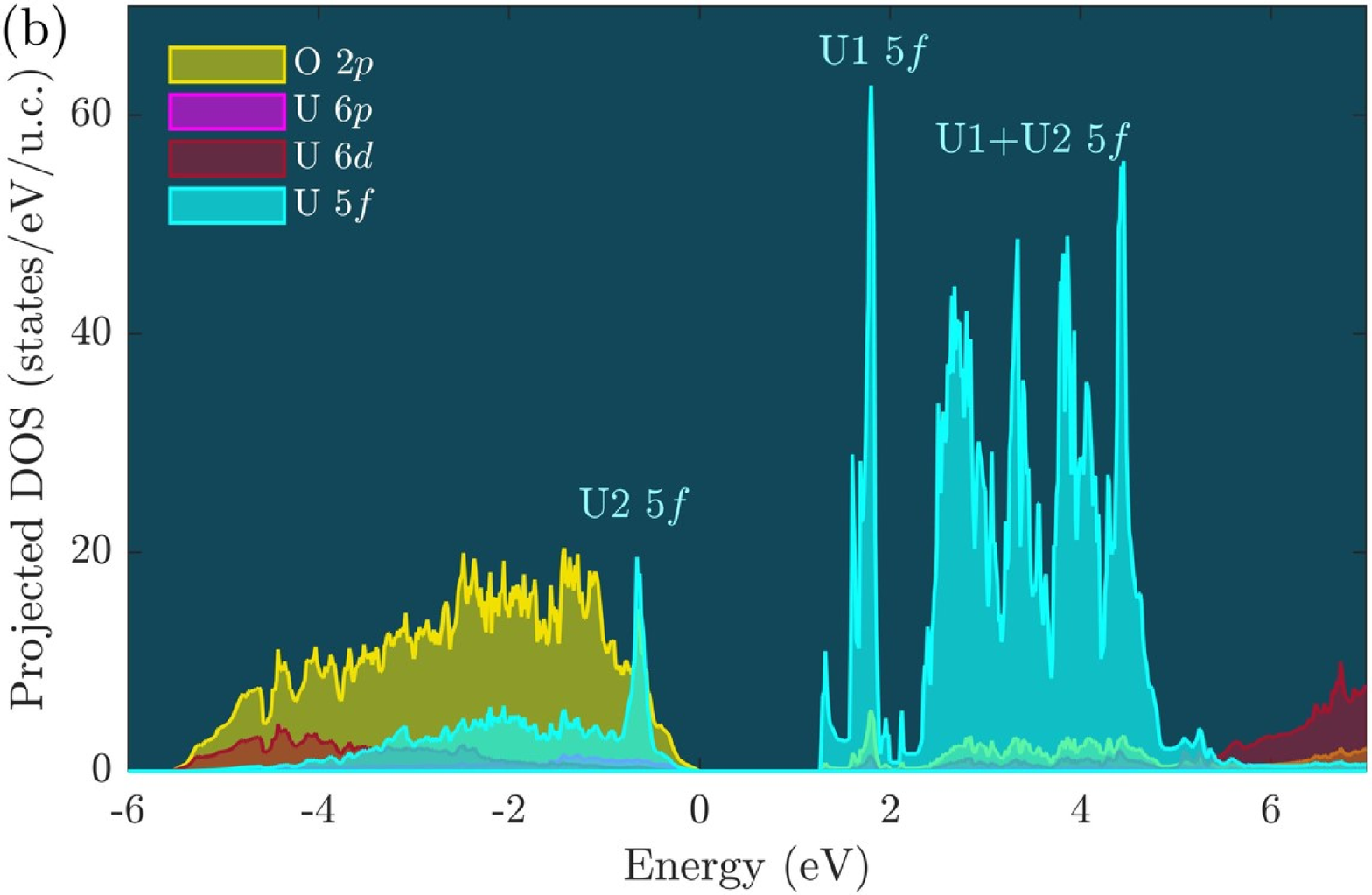}
\caption{\label{fig3} (a)
Band structure of U$_3$O$_8$, with the colors highlighting
the contributions of the U $5f$, O $2p$, and U $6d$ states, which are the main contributors in the energy range shown [the valence band maximum (VBM) is at 0]. Close inspection, using
a $10\times10\times14$ {\bf k}-point mesh, shows that the conduction band minimum falls
within the triangle formed by the $\Gamma$, X, and Y (see Fig.~\ref{fig4}).
(b) Atom-type projected density of states. The density of states calculations indicate that the band gap is
close to 1.27 eV. The U1 and U2 $5f$ labels are for reference in the main text.}
\end{figure}

\begin{figure}[!h]
\includegraphics[width=0.6\hsize]{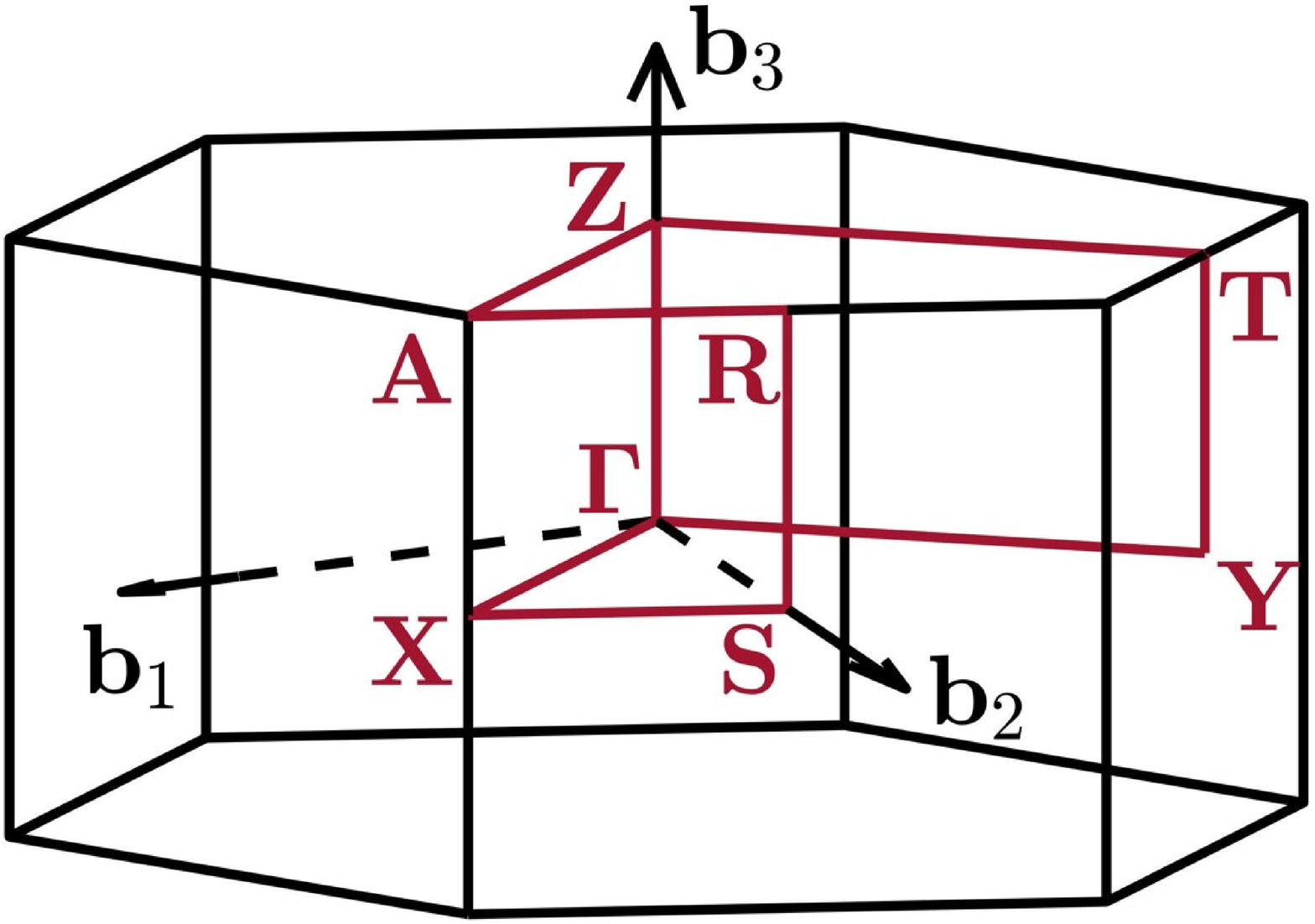}
\caption{\label{fig4} 
{\bf k}-point path for the band plot in Fig.~\ref{fig2}(a). Note here we
use the $C2mm$ setting of the orthorhombic
cell (see Fig.~12 in Ref.~\citenum{setyawan2010}).}
\end{figure}

\clearpage

\begin{center}
\subsection{IV. Projected densities of states}
\end{center}

\begin{figure}[!h]
\begin{center}
\includegraphics[width=0.36\hsize]{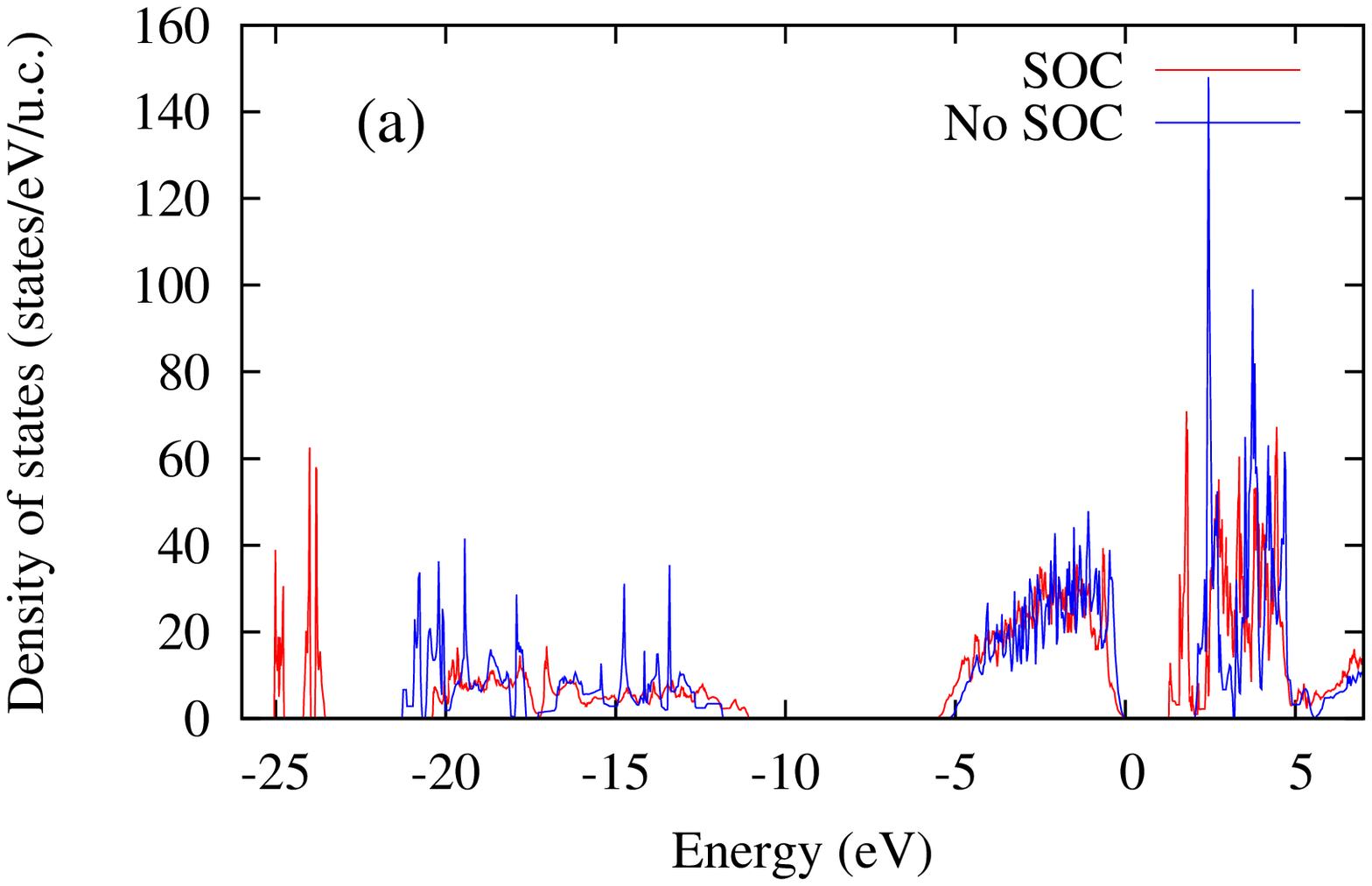}
\includegraphics[width=0.36\hsize]{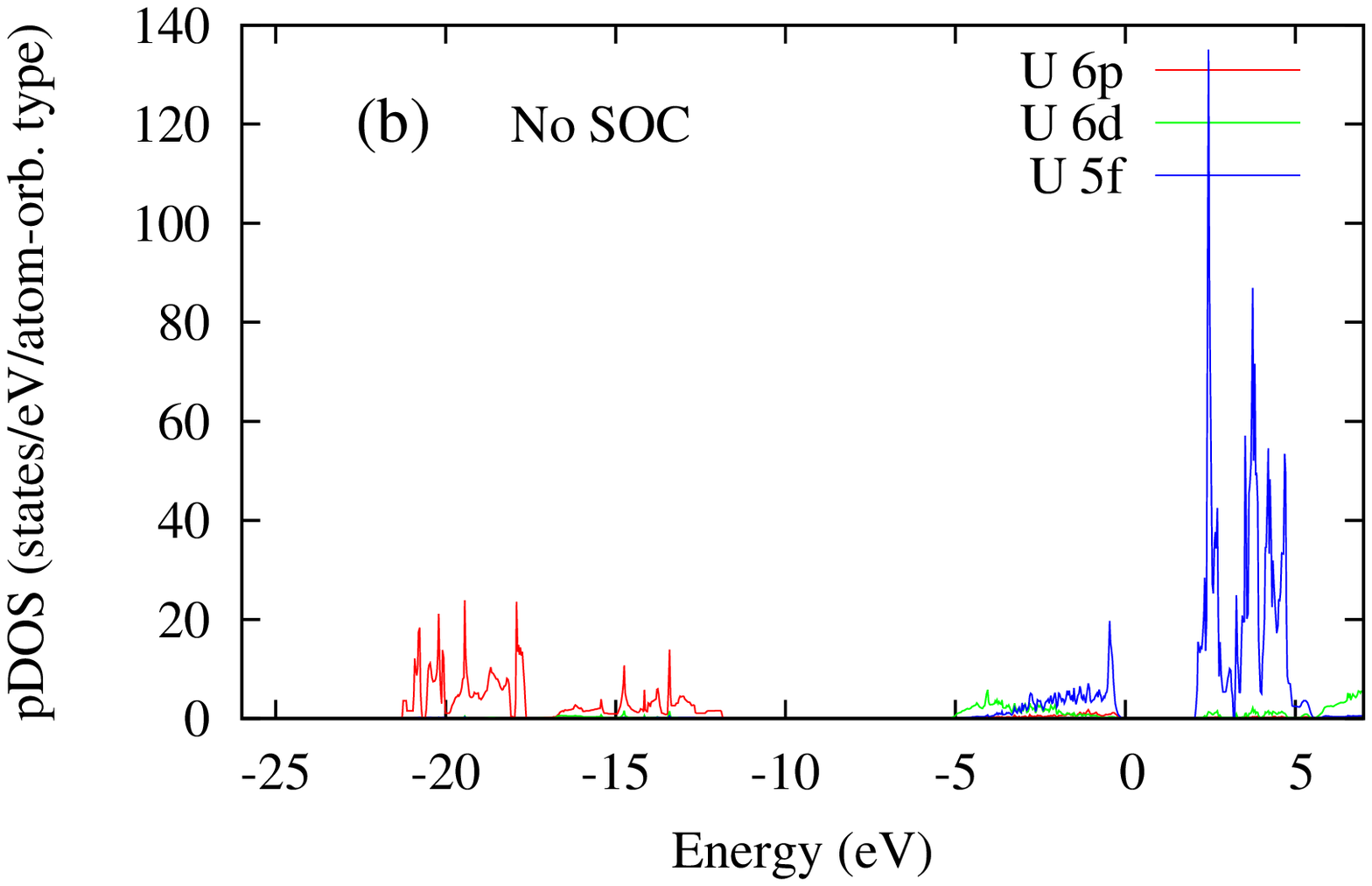}
\includegraphics[width=0.36\hsize]{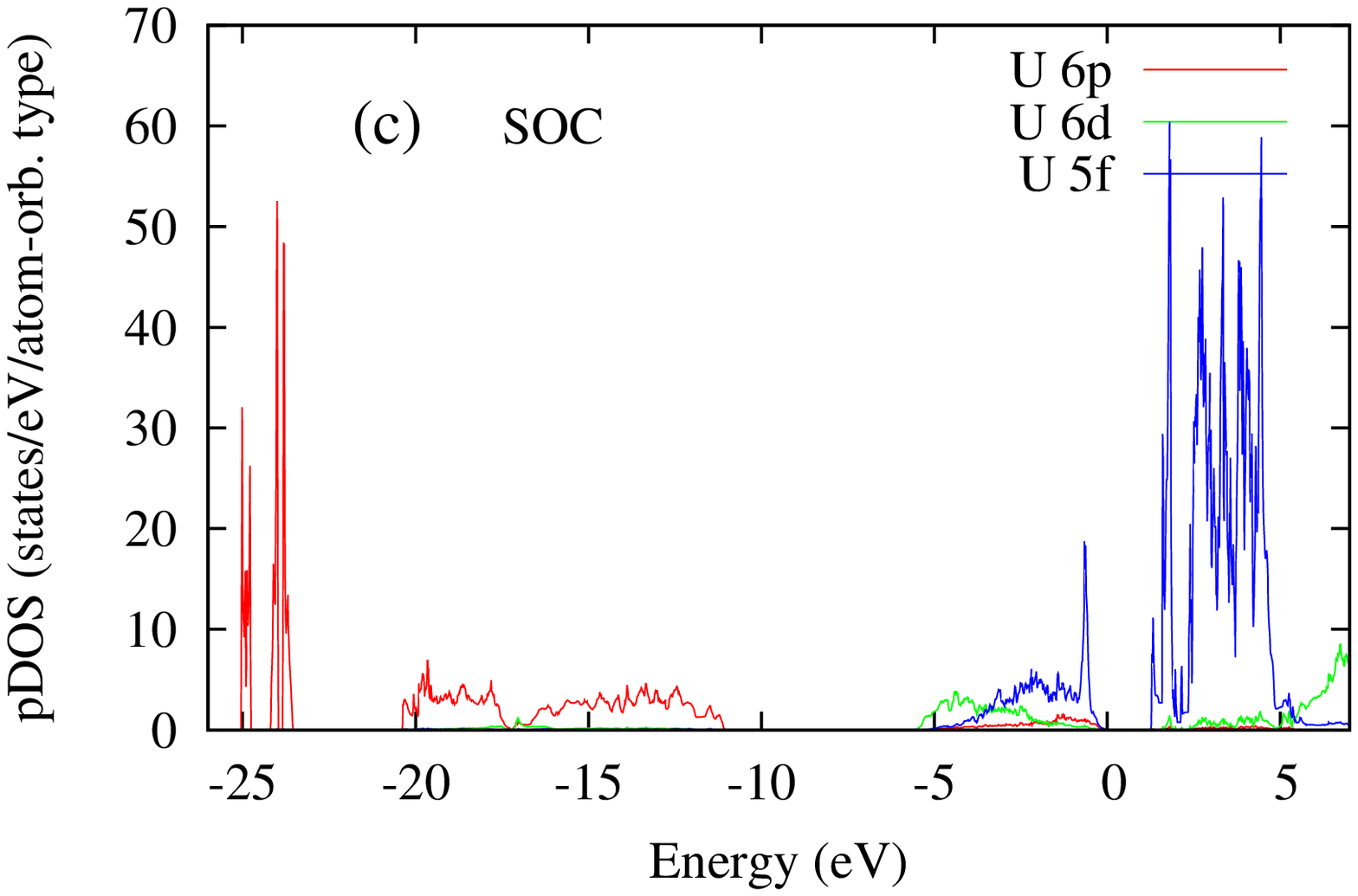}
\includegraphics[width=0.36\hsize]{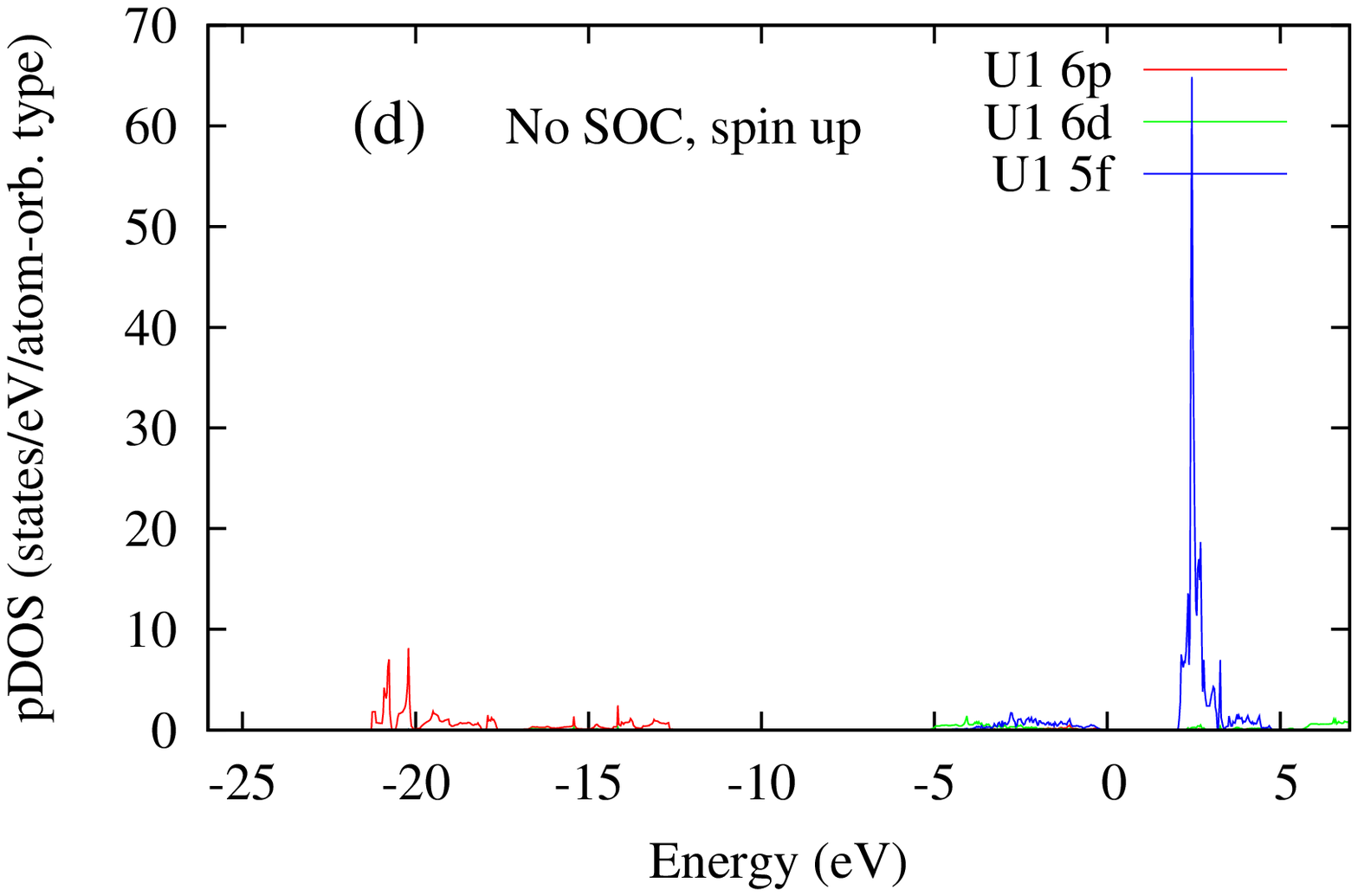}
\includegraphics[width=0.36\hsize]{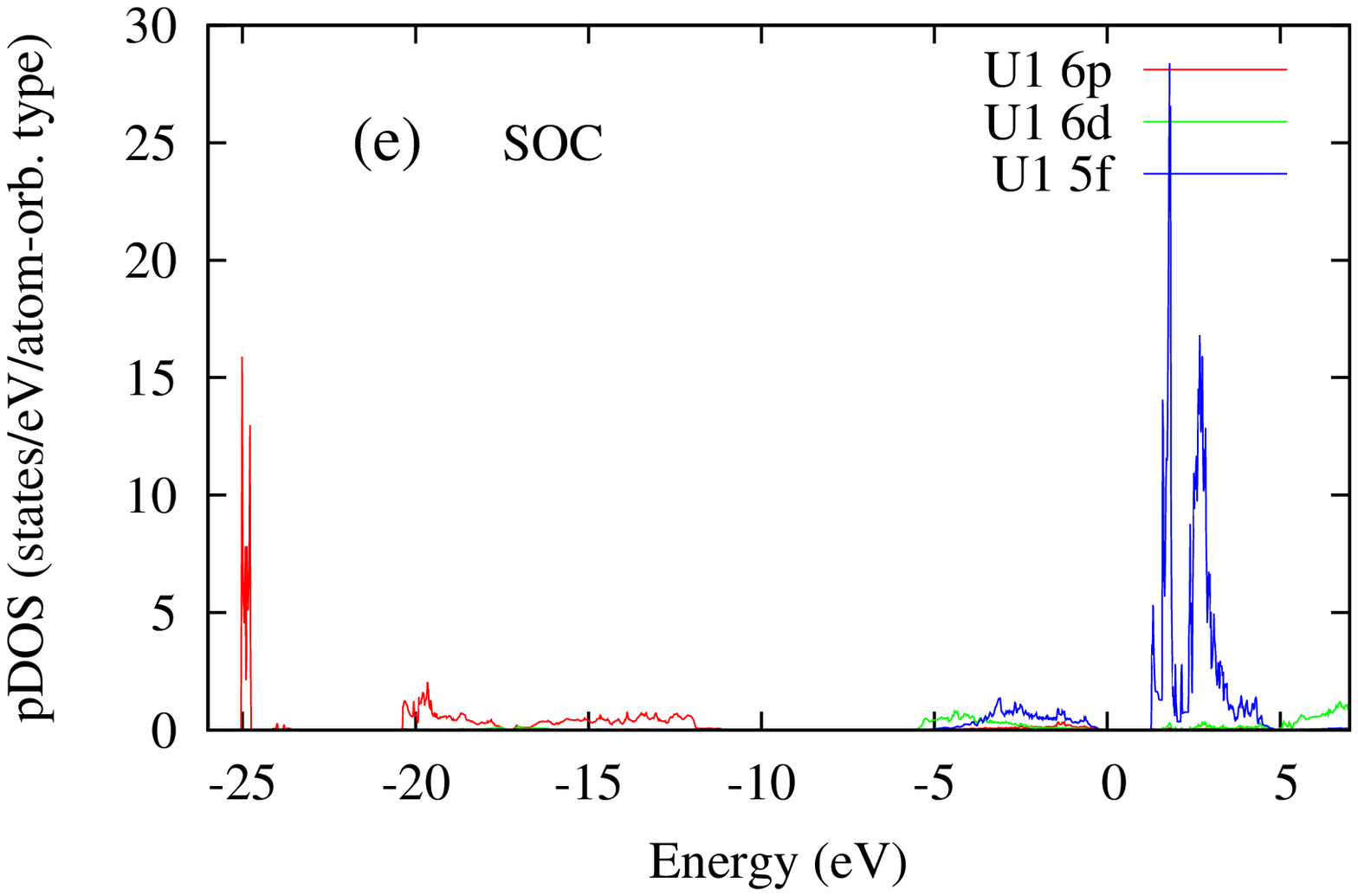}
\includegraphics[width=0.36\hsize]{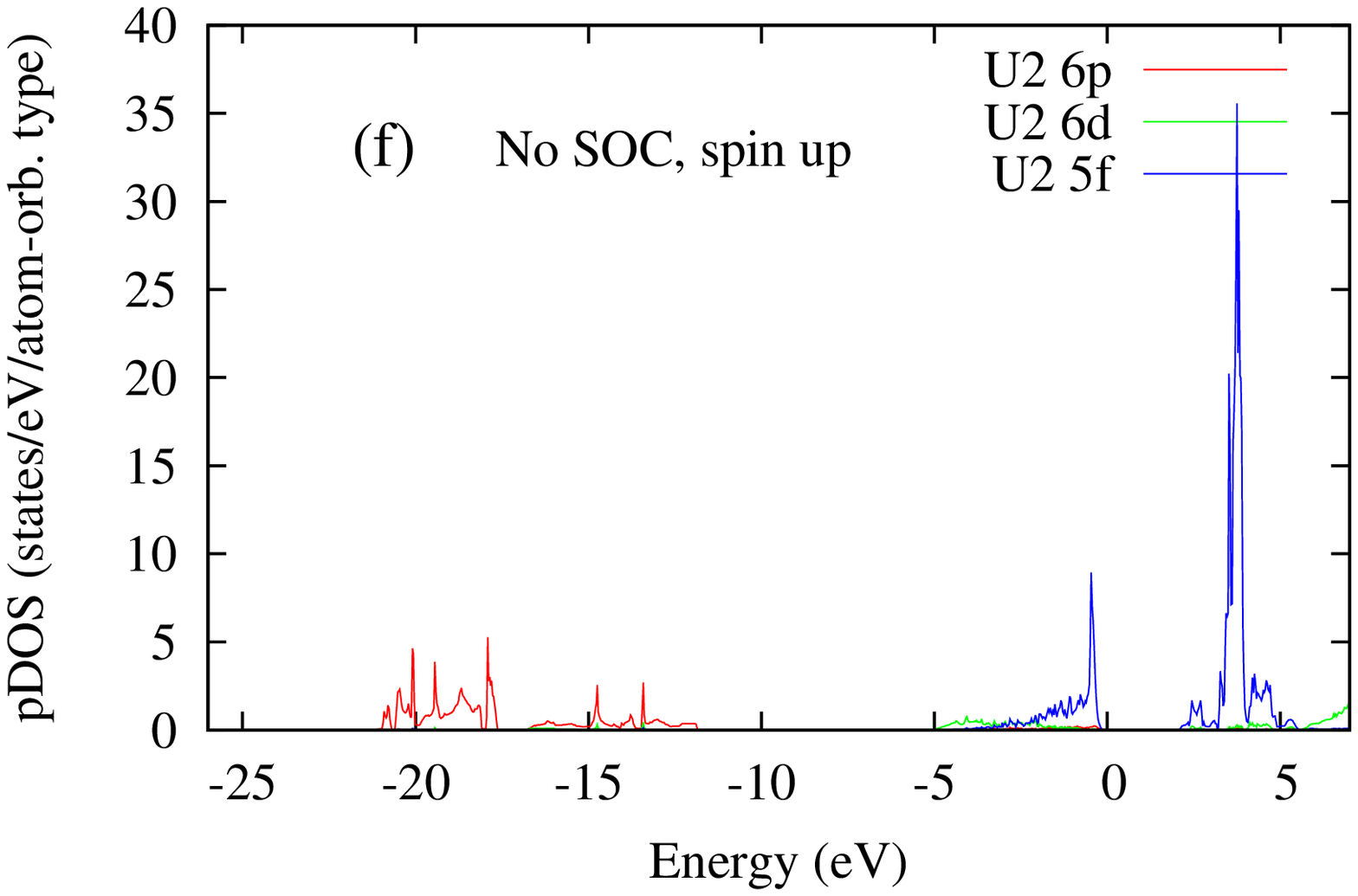}
\includegraphics[width=0.36\hsize]{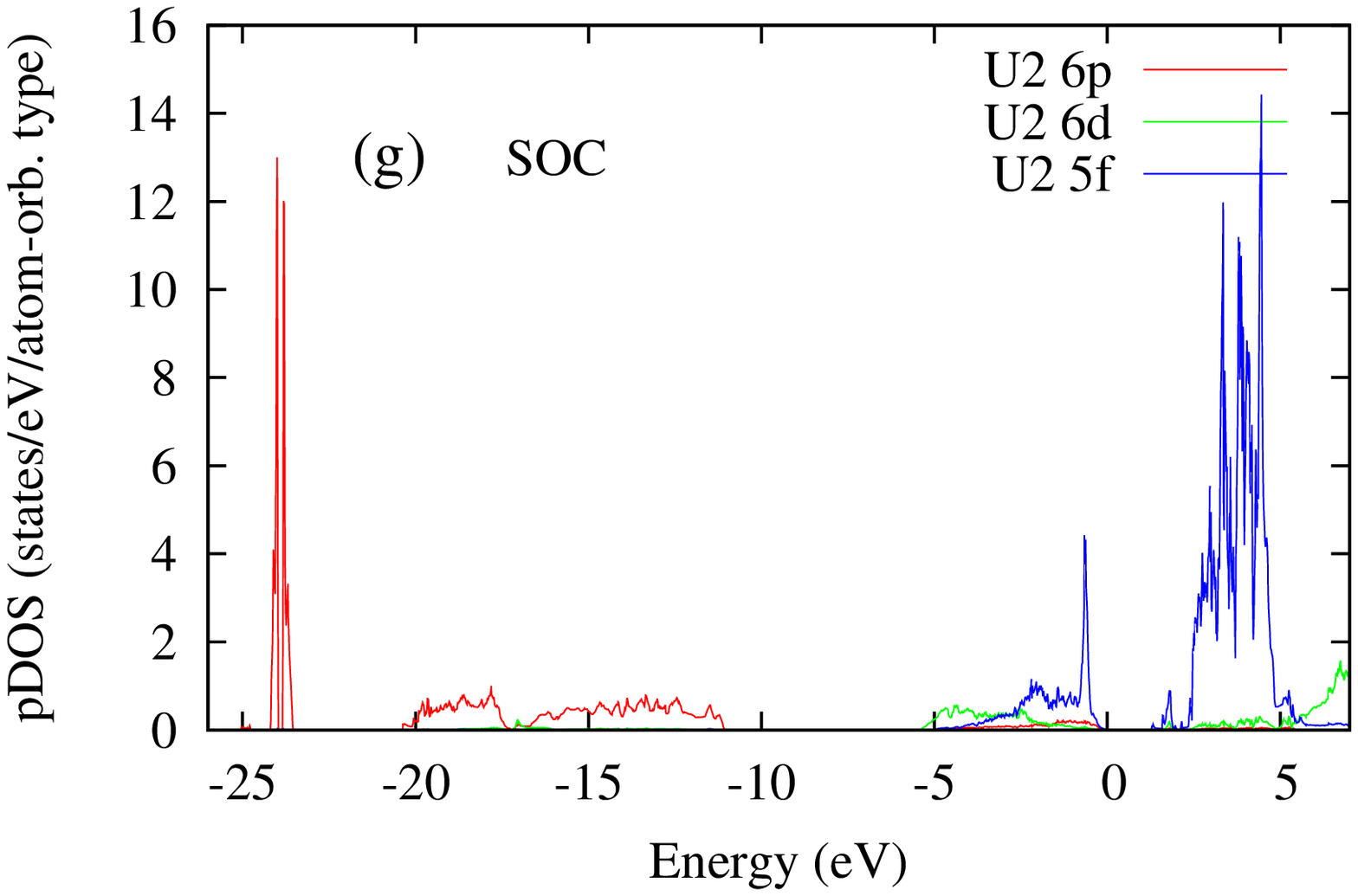}
\includegraphics[width=0.36\hsize]{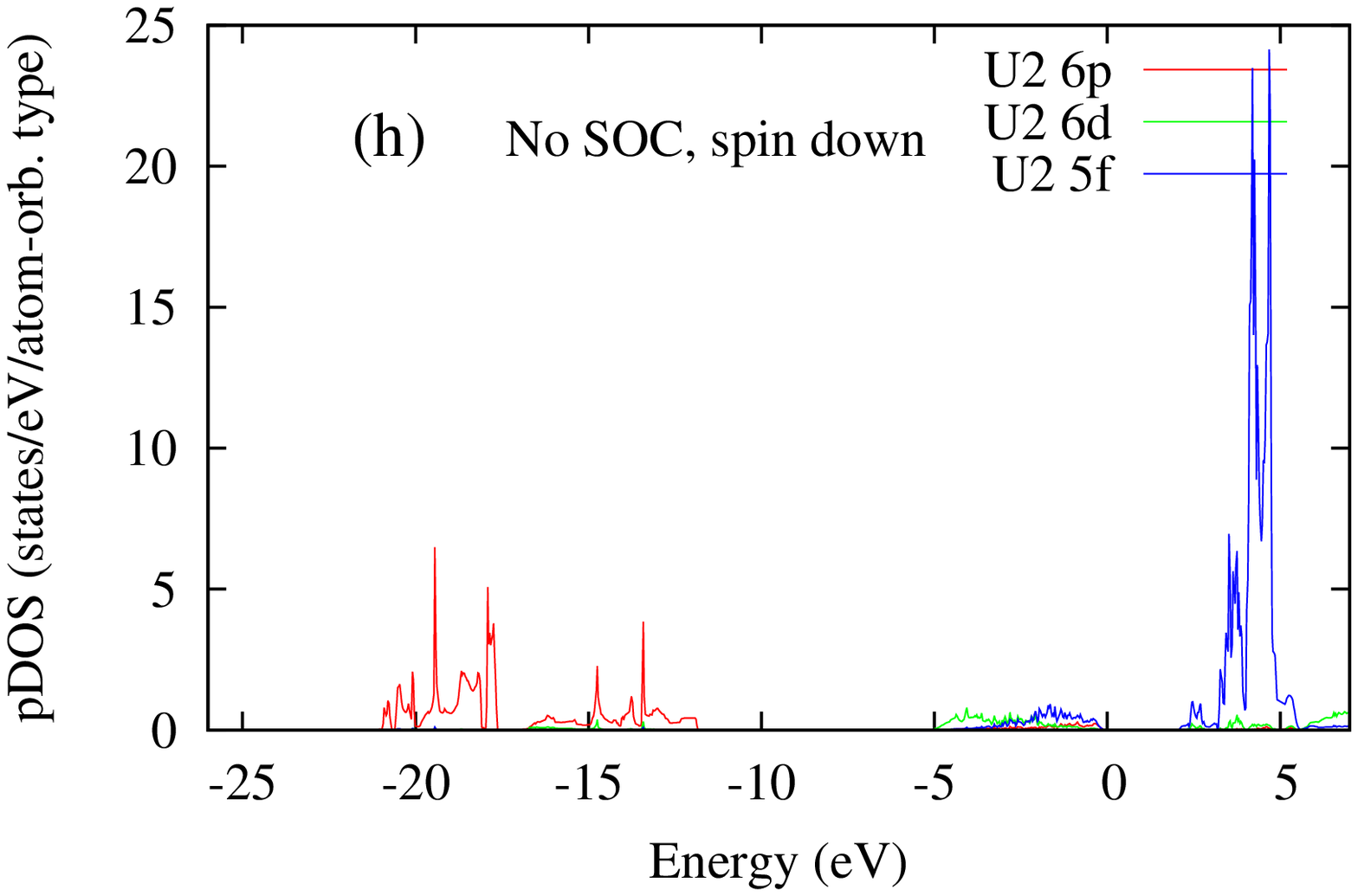}
\end{center}
\caption{\label{fig5} (a) Comparison of DOS with and without SOC over a wider range of
energies than in the main text (VBM is at 0). SOC increases the band widths, as can be seen in
the conduction band, and especially in the lower valence band, where there is a split not
present in the calculations neglecting SOC. Which states make the difference at which energies
 is made clear in (b) and (c). More specifically, figures (d) and (e) compare the contributions
 of uranium atom U1, which carry no net magnetic moment. Clearly, SOC lowers the energy of
 the $5f$ states with respect to the VBM, while increasing the band width. Further, with SOC the
 $6p$ states split, with a narrow, strong peak at $\sim-25$ eV. (The DOS of U1 for down spin is
 indistinguishable from the up spin DOS, as the magnetic moment is 0). Figures (f), (g), and (h)
 compare the contribution of the uranium U2 atoms, which carry a net magnetic moment. Figures
 (f) and (h) show that the magnetization is essentially due to the $5f$ states just below the VBM.
 Figure (g) again shows that SOC causes a splitting of the $6p$ states, with a narrow, strong
 peak below a shallower extended continuum. These $6p$ states (from U1 and U2) contribute to
 lower considerably the total energy of the system.}
\end{figure}

\clearpage

\begin{center}
\subsection{V. Dielectric function}
\end{center}

\begin{figure}[!h]
\begin{center}
\includegraphics[width=0.49\hsize]{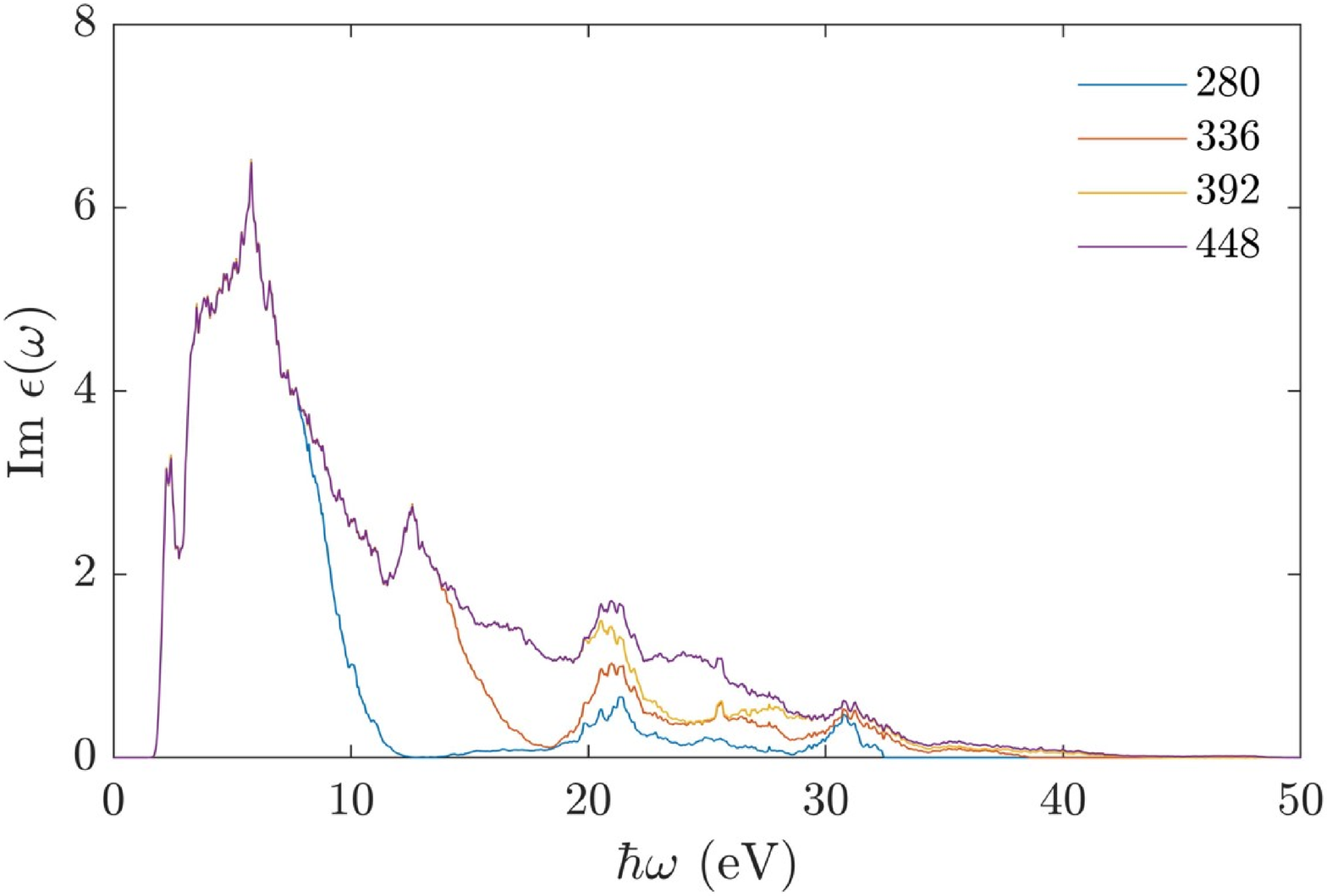}
\includegraphics[width=0.49\hsize]{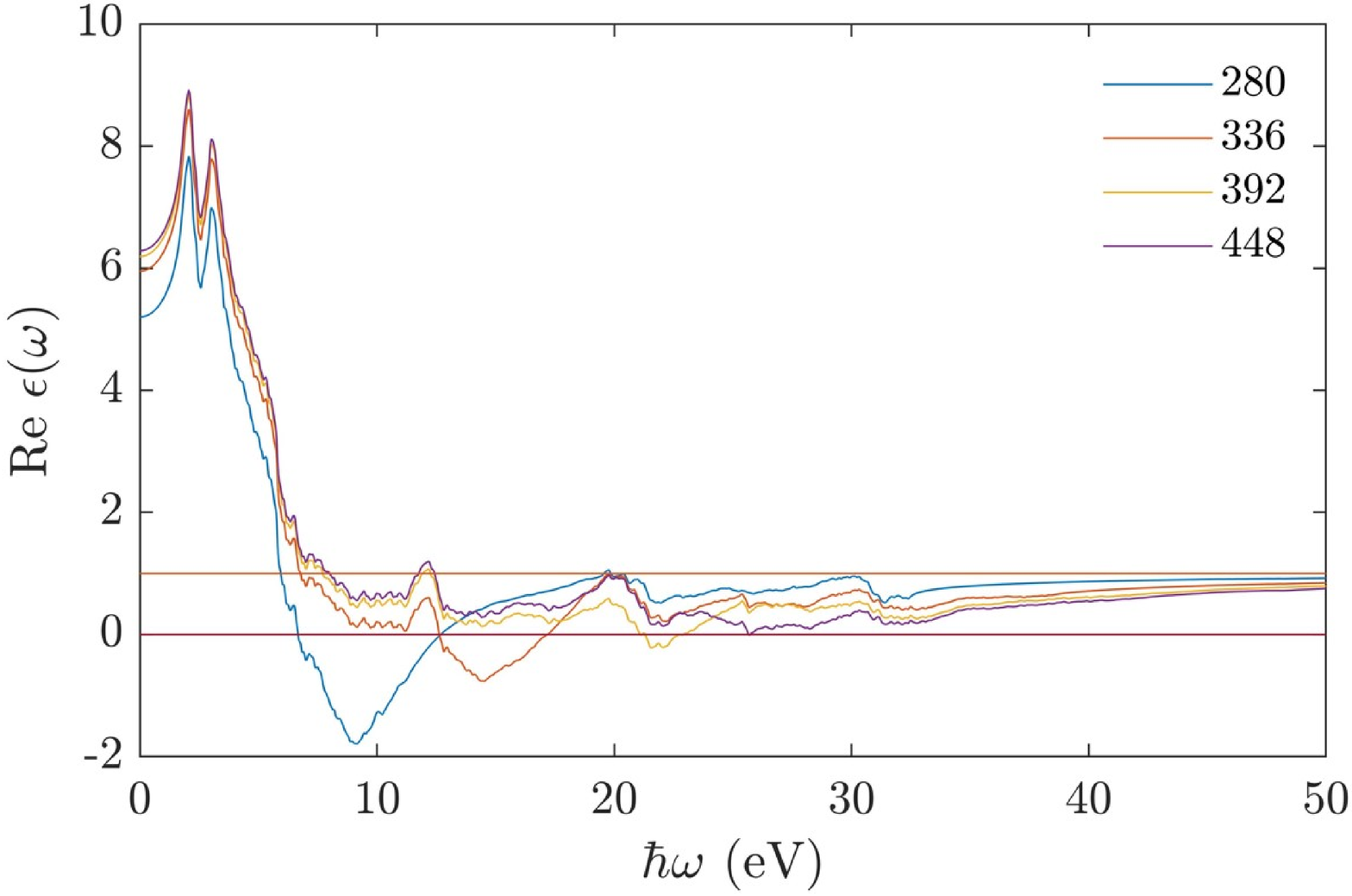}  
\end{center}
\caption{\label{fig6} Imaginary part (left panel) and real part (right panel) of the
dielectric function. Convergence of the real part is more difficult because it is calculated
based on the imaginary part, using the Kramers-Kronig relations. For the real part to be converged up to a certain energy, sufficient bands should be included. Here convergence is reasonable up to
5--6 eV with 448 bands.}
\end{figure}